\documentclass{article}
\usepackage{LaThuileFPSpro}
\usepackage{graphicx}
\newcommand{\qqbar}{q\bar{q}}
\newcommand{\bbbar}{b\bar{b}}
\newcommand{\ppbar}{p\bar{p}}
\newcommand{\ttbar}{t\bar{t}}
\newcommand{\BR}{\mbox{BR}}

\begin{document}
\title{ 
  STANDARD MODEL HIGGS SEARCHES AND PERSPECTIVES AT THE
  TEVATRON
  }
\author{
  Stefan S\"oldner-Rembold\\
  {\em University of Manchester, Oxford Road, Manchester, M13 9PL,
  United Kingdom} \\
  On behalf of the CDF and D\O\ Collaborations\\
  }
\maketitle

\baselineskip=11.6pt

\begin{abstract}
The status and perspectives of Standard Model Higgs searches
at the Tevatron experiments CDF and D\O\ are discussed.
\end{abstract}
\newpage
\section{Introduction}
\begin{figure}[htbp] 
\begin{center} 
\includegraphics[width=0.45\textwidth]{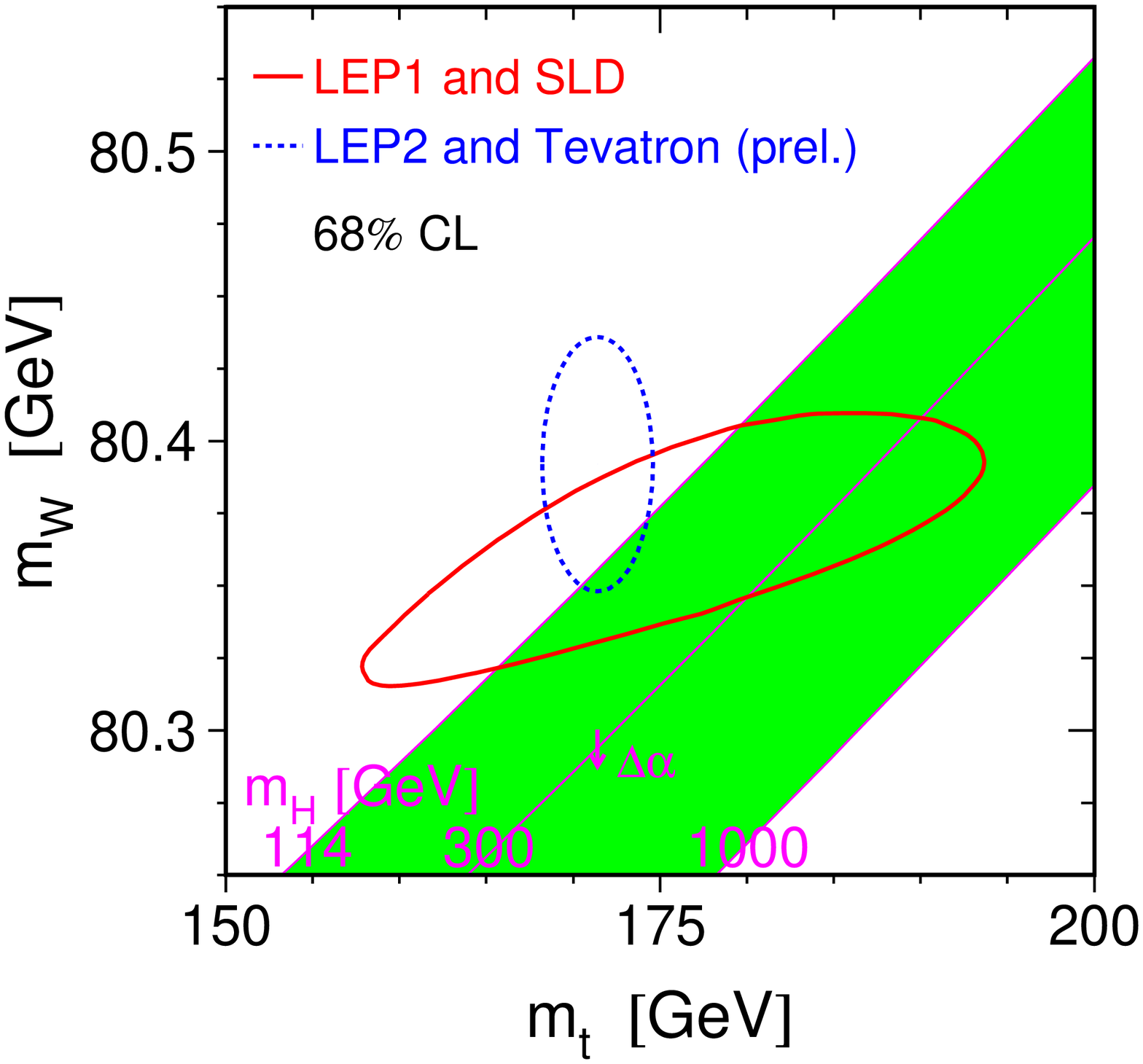} 
\includegraphics[width=0.45\textwidth]{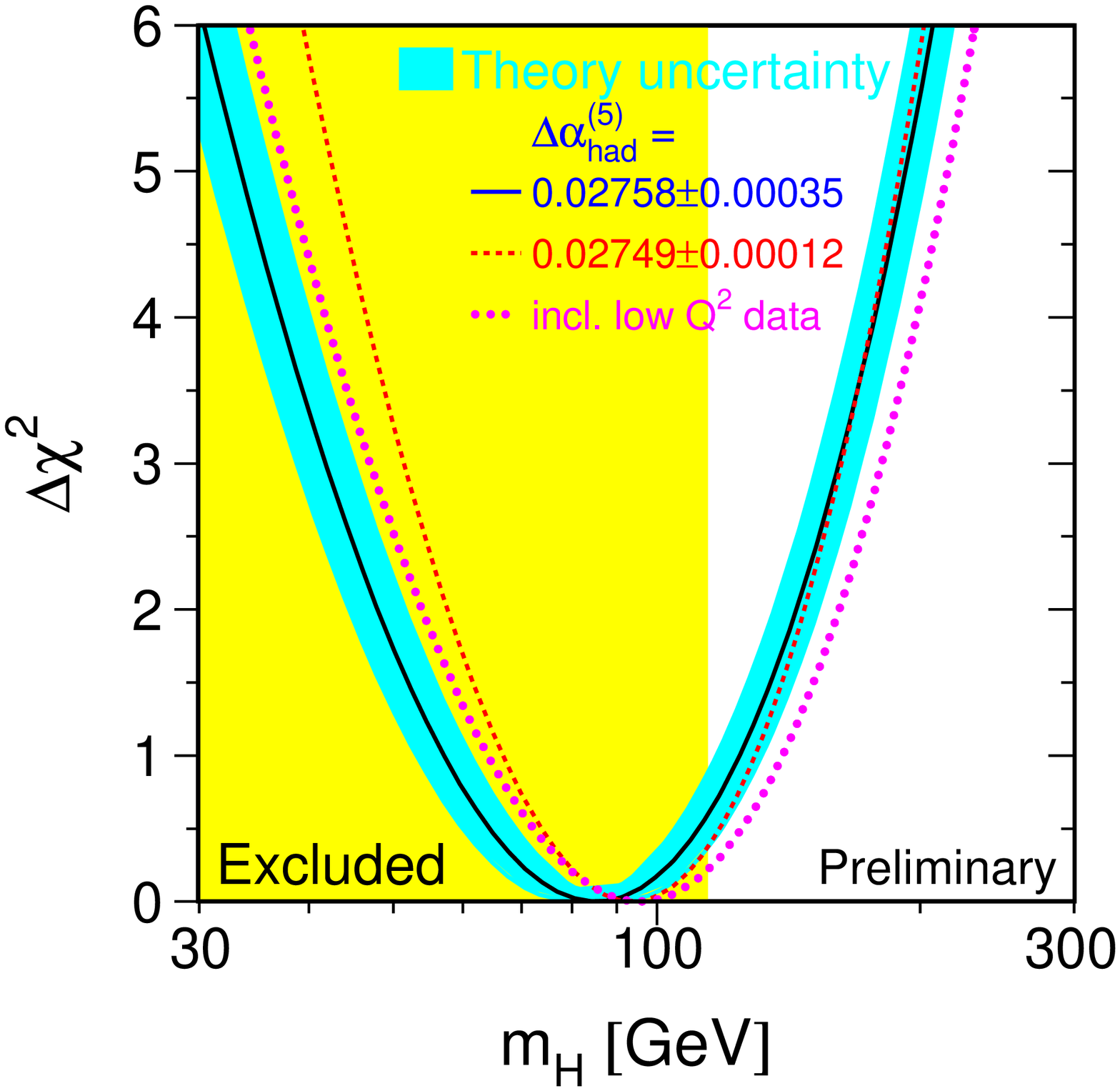} 
\end{center} 
\caption[]{
Dependence of the Higgs
mass on the measured $W$ and top masses (left).
$\Delta\chi^2$ curve derived from precision electroweak data
as a function of Higgs mass (right)~\protect\cite{bib-lep}.
}
\label{fig1} 
\end{figure}

In the Standard Model (SM) the Higgs mechanism is responsible
for breaking the electroweak symmetry, thereby giving mass to the
the $W$ and $Z$ bosons. It predicts the existence of a heavy scalar boson, the
Higgs boson, with a mass that can not be predicted by the SM. 
The Tevatron experiments, D\O\ and CDF, constrain the mass of the
SM Higgs bosons indirectly through electroweak precision
measurements. The main contribution to these indirect constraints
from the Tevatron are the measurements of the $W$ and top 
masses~\cite{bib-barberis}. The dependence of the Higgs
mass on these measurements is shown in Figure~\ref{fig1} together
with the Higgs mass dependence on the measured 
electroweak precision
parameters. The SM fit yields a best value of
$m_H=85^{+39}_{-28}$~GeV~\cite{bib-lep}. 

The direct mass
limit from LEP is
114.4~GeV at $95\%$ Confidence Level (CL)\cite{bib-higgs}. 
The Tevatron experiments
search for direct Higgs boson production in the mass range
above the LEP limit using associated Higgs production 
($\ppbar\to WH$, $\ppbar\to ZH$) and gluon fusion ($gg\to H$). 
The main tools are jet
reconstruction, $b$ tagging and lepton identification. 
The tagging of $b$ jets from Higgs decays is usually performed
by requiring tracks with large impact parameters or by
reconstructing secondary vertices. More advanced analyses
will combine these and other variables in Neural Nets.
Events with neutrinos in the final state are identified
using missing transverse energy. The reconstruction
of all these variables require excellent performance
of all detector components.

There are different types of background to the Higgs search.
An important source of background are
multi-jet events (QCD background). This background
and the instrumental
background due to mis-identified leptons or $b$-jets is not
very well simulated by Monte Carlo. Its shape and normalisation
is therefore taken from data using control samples.
Electroweak background processes such as $VV (V=W,Z)$ 
or $\ttbar$ pair production often dominate at the final stages
of the selection; these are simulated using Monte Carlo
programs.

At the time of writing these proceedings, the two Tevatron experiments
have each recorded about 1.5 fb$^{-1}$ of luminosity. Most results
presented here are based on data corresponding to about 0.3 fb$^{-1}$.
The preliminary results of the two collaborations are
accessible through their web pages~\cite{bib-web}.

\section{Low Mass Higgs Searches}
In the Higgs mass region below 135 GeV the most important search
channels are the associated production of a Higgs and
a $W$ or $Z$ boson. 
The largest
production cross-section is due to the process $gg\to H$, it is
about $0.7$~pb for a Higgs mass of 115 GeV. 
However, the Higgs boson
predominantly decays into $b$ quarks in this mass range and the 
QCD background is too large in this final state.
The process $gg\to H$ is therefore not a viable
search channel. The $WH$ and $ZH$ channels with the vector boson
decaying into leptons have much lower cross-sections
but the lepton tag and cuts on missing transverse energy
help to reduce the background significantly.

\subsection{$ WH \to \ell \nu \bbbar$}

This final state consists of two $b$ jets from the Higgs boson and
a charged lepton (electron or muon) and a neutrino from the $W$ boson.
D\O\ therefore selects events with one or two tagged $b$-jets, one isolated
lepton with a transverse momentum $p_T>20$~GeV and 
missing transverse energy $E_T^{\rm miss}>25$ GeV. 
The main backgrounds are $W$ production in association with
two heavy flavour jets and $\ttbar$ production. This search
yields an upper limit on the $WH$ production cross-section in
the range 2.4 pb to 2.9 pb for Higgs masses between 105 GeV and 145 GeV.
The average integrated luminosity is 378 pb$^{-1}$.

CDF performed a similar analysis 
requiring one or two tagged b-jets with $E_T>15$~GeV, 
an electron or a muon with $p_T>20$ GeV and 
$E_T^{\rm miss}>20$~GeV~\cite{bib-whcdf}.
This search 
has recently been updated to include data with
an integrated luminosity of $955$~pb$^{-1}$, yielding limits
on the cross-section $\sigma(\ppbar\to WH) \BR (H \to \bbbar)$ 
between 3.9~pb and 1.3~pb for the mass range 110~GeV to 
150~GeV~\cite{bib-dis}.

\subsection{$ZH \to \nu\nu \bbbar$}
This channel has
very good sensitivity because of the large branching ratios
for $Z\to\nu\nu$ and $H\to\bbbar$ decays. Since the two b-jets
are boosted in the transverse direction, the signature for
the final state are acoplanar di-jets in contrast to 
most QCD di-jet events which are expected to be back-to-back in
the transverse plane. In addition, large missing transverse
energy (D\O\ : $E_T^{\rm miss}>50$~GeV, CDF: $E_T^{\rm miss}>75$~GeV) 
is required. 
Main background sources are $W/Z$ production in association
with heavy flavour jets, QCD events and $\ttbar$ pairs. 
A search in 260 pb$^{-1}$ of data has been published by
D\O\ \cite{bib-zhd0} giving a limit on 
$\sigma(\ppbar\to ZH) \BR (H \to \bbbar)$ of 3.4-2.5~pb
for masses 105-135~GeV.
The most recent preliminary CDF result
includes 973 pb$^{-1}$ of data~\cite{bib-dis}.

\subsection{$ZH \to \ell\ell \bbbar$}
In this channel the $Z$ boson is reconstructed by its decay into 
two high-$p_T$ isolated muons or electrons. The reconstructed $Z$ 
and two b-tagged
jets are then used to select the Higgs signal. The main background sources
are $Z$ plus heavy jets and $\ttbar$ production. 
D\O\ measured a limit on  
$\sigma(\ppbar\to ZH) \BR (H \to \bbbar)$ 
between 6.1~pb and 4.7~pb in the Higgs
mass range 115-145~GeV using 320-389~pb$^{-1}$ of data.
CDF performed a similar analysis using about 1 fb$^{-1}$ of data 
and obtained a limit of 2.0-1.1~pb in the mass range 
110-150~GeV~\cite{bib-dis}.

\section{High Mass Higgs Searches}
The dominant decay mode for higher Higgs masses is $H\to WW^{(*)}$.
Leptonic decays of the W bosons are therefore used to suppress QCD background.
\subsection{$H \to WW^{(*)}  \to \ell\ell$}
The signature of the $gg\to H \to WW^{(*)}$ channel is two high-$p_T$
isolated leptons with small azimuthal separation, $\Delta\phi_{\ell\ell}$,
due to the spin-correlation between the final-state leptons in
the decay of the spin-0 Higgs boson.
In contrast, the lepton pairs from $Z$ decays are predominantly
back-to-back in $\phi_{\ell\ell}$. The experiments currently
use the $W$ decays into electrons or muons for this search. 
The CDF limit on 
$\sigma(\ppbar\to W) \BR (H \to WW)$ using 360~pb$^{-1}$ is between
5.5~pb and 3.2~pb for Higgs masses from 120 to 200~GeV~\cite{bib-wwcdf}.
D\O\ has recently updated a published analysis with 
300-325 pb$^{-1}$~\cite{bib-wwd0} using 950 pb$^{-1}$. The 
resulting limits are 6.3-1.9~pb for Higgs masses in the range
120-180~GeV~\cite{bib-dis}.

\subsection{$H \to WWW^{*}  \to \ell \nu \ell^{'} \nu\qqbar$}

Here the Higgs boson is produced in association with a $W$ boson.
D\O\ has performed an analysis with data corresponding to
360-380~pb$^{-1}$, requiring two opposite sign isolated leptons ($e,\mu$)
with $p_T>15$~GeV, one originating from the $ H \to WW$ decay and one from the
associated $W$ boson, and $E_T^{\rm miss}>20$~GeV~\cite{bib-wwwd0}.
This opposite sign charge requirement is very powerful in rejecting
background from $Z$ production. The remaining background is either
due to di-boson production or due to charge mis-measurements.
The upper limit on the cross-section $\sigma(\ppbar\to W) \BR(H\to WW^*)$
is between 3.2~pb and 2.8~pb for Higgs masses from 115~GeV to 175~GeV. 
A similar analysis was performed by CDF with an integrated luminosity
of 194~pb$^{-1}$.
\section{Combined Tevatron Limit}
The data of both experiments have been combined using the full
set of analyses with luminosities in the range 
0.2-1~fb$^{-1}$~\cite{bib-comb}. The $95\%$ Confidence Level (CL) 
upper limits are a factor of 10.4 (3.8) higher than the expected
SM cross-sections for Higgs masses of 115 (160) GeV. 
\begin{figure}[htbp] 
\begin{center} 
\includegraphics[width=0.8\textwidth]{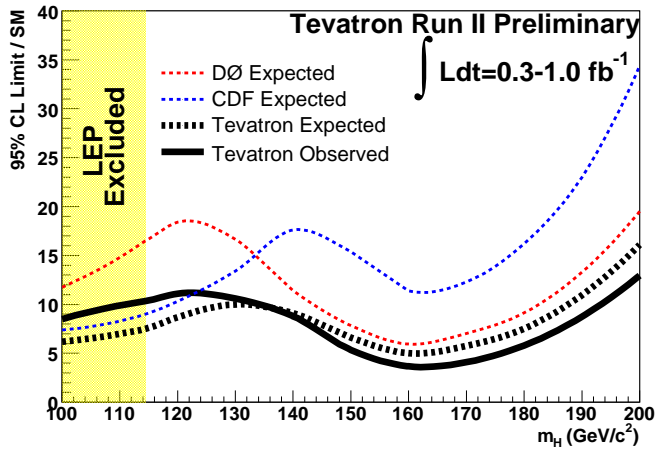} 
\end{center} 
\caption[]{
Expected and observed $95\%$ CL cross-section ratios for the combined
CDF and D\O\ analyses. The expected $95\%$ CL ratio for both experiments
are also shown.
}
\label{fig-comb} 
\end{figure} 
The difference between the expected limits for CDF and D\O\ is
mainly due to the different size data samples that have been analysed in
the low and high mass high region, The overall sensitivity
shown is therefore closer to the sensitivity of a single
experiment with 1 fb$^{-1}$.
\section{Summary and Perspectives}
The CDF and D\O\ experiments at Fermilab's Tevatron have searched
for the SM Higgs boson in a variety of channel, using data 
corresponding to integrated luminosities between 0.2~fb$^{-1}$
 and 1~fb$^{-1}$.
The cross-section limits are about a factor ten above the SM
expectation for a Higgs boson mass of 115 GeV. A long list of
improvements - in addition to adding more data - are expected to
lead to a significant increase in the sensitivity. The most
important improvements are
better $b$ tagging using Neural Net algorithms, improvements of the 
di-jet mass resolution to separate the Higgs signal from the background,
the inclusion of new channels such as $WH\to\tau(\to \mbox{hadrons})\nu
\bbbar$ and the use of advanced analysis techniques such as Neural Nets. 
This and more data will give the Tevatron experiments
a good chance to exclude a SM Higgs boson (or find indications 
for its existence) in the low mass region above the LEP limit of 114.4~GeV.
\section*{Acknowledgements}
The author would like to thank the organisers for making this 
a very enjoyable conference and the Royal Society for 
the conference grant.


\begin{thebibliography}{99}
\bibitem{bib-barberis}
E. Barberis, Fermilab-Conf-060146, these proceedings.
\bibitem{bib-lep}
LEP Electroweak Working Group, 
\mbox{http://lepewwg.web.cern.ch/LEPEWWG/}
\bibitem{bib-higgs}
ALEPH, DELPHI, L3 and OPAL Collaborations, Phys. Lett. {\bf B} 565, 61 (2003).
\bibitem{bib-web}
\mbox{http://www-cdf.fnal.gov/physics/exotic/exotic.html}
\mbox{http://www-d0.fnal.gov/Run2Physics/WWW/results/higgs.htm}
\bibitem{bib-whcdf}
CDF Collaboration, Phys. Rev. Lett {\bf 96}, 081803 (2006).
\bibitem{bib-dis}
This result was presented after the conference.
\bibitem{bib-zhd0}
D\O\ Collaboration, hep-ex/0607022, submitted to Phys. Rev. Lett.
\bibitem{bib-wwcdf}
CDF Collaboration, Phys. Rev. Lett {\bf 97}, 081802 (2006).
\bibitem{bib-wwd0}
D\O\ Collaboration, Phys. Rev. Lett {\bf 96}, 011801 (2006).
\bibitem{bib-wwwd0}
D\O\ Collaboration, hep-ex/0607032, submitted to Phys. Rev. Lett.
\bibitem{bib-comb}
D\O\ and CDF Collaborations, CDF Note 8384 and D\O\ Note 5227.
\end{thebibliography}
\end{document}